\documentstyle[12pt]{article}

\begin{document}

\title{Quadratic reheating}

\author{Luis P. Chimento and Alejandro S. Jakubi      \\
{\it Departamento de F\'{\i}sica,  }\\
{\it  Facultad de Ciencias Exactas y Naturales, }\\
{\it Universidad de Buenos Aires }\\
{\it  Ciudad  Universitaria,  Pabell\'{o}n  I, }\\
{\it 1428 Buenos Aires, Argentina.}}

\maketitle

\begin{abstract}

The reheating process for the inflationary scenario is investigated
phenomenologically. The decay of the oscillating massive inflaton field into
light bosons is modeled after an out of equilibrium mixture of interacting
fluids within the framework of irreversible thermodynamics. Self-consistent,
analytic results for the evolution of the main macroscopic magnitudes like
temperature and particle number densities are obtained. The models for linear
and quadratic decay rates are investigated in the quasiperfect regime. The
linear model is shown to reheat very slowly while the quadratic one is shown
to yield explosive particle and entropy production. The maximum reheating
temperature is reached much faster and its magnitude is comparable with the
inflaton mass.

\end{abstract}

\vskip 3cm

\noindent
PACS 98.80.Hw, 04.40.Nr, 05.70.Ln

\newpage
\section{Introduction}

During an inflationary stage driven by the potential energy of a classical
scalar field, the temperature of the universe was redshifted to almost zero.
Thus a mechanism to raise the temperature of the universe at the end of the
inflationary stage is required to match this scenario with the standard hot
Big Bang cosmology \cite{Kolb90,Lin90}.

Currently it is assumed that soon after the end of inflation, the scalar field
began to oscillate near the minimum of its effective potential. Decay of this
scalar field through particle production and thermalization of its energy with
a huge increase in temperature and large entropy production, occurred during
the reheating period. It started with violent particle production through
parametric resonance, also called the preheating process, and lead to a highly
nonequilibrium distribution of the produced particles, subsequently relaxing
to an equilibrium state \cite{KoLiSta1,Boya1,Shta}.

In this paper we will consider the decay of the massive homogeneous inflaton
scalar field $\phi$ coupled with light bosons $\chi$, as it arises in chaotic
inflationary models. A detailed quantum mechanical analysis is quite
complicated and involves taking into account nonlinear processes, backreaction
effects, rescattering, etc \cite{Lin97}. Further, since particles created
during the preheating stage are far from equilibrium, their thermalization is
achieved via collisional relaxation. This process of thermalization has not
been understood yet from first principles \cite{Hu97a}. Here,
we follow an alternative description
of the physics, and instead of both fields and
their interaction we consider an effective, phenomenological model in
terms of an out-of-equilibrium mixture of two reacting fluids within the
framework of relativistic irreversible thermodynamics \cite{Isr76,Ger90}. The
difference in the equations of state of the fluid components and the
simultaneous deviation from the detailed balance in the rate equations for the
interfluid reactions  give rise to an entropy production effect that manifests
itself as a reactive bulk pressure \cite{Zim97b}. This two-fluid model can be
described in the effective one-fluid picture, based on a single temperature
that we identify with the temperature of the reheating process. One of the
advantages of this approach is that it is possible to treat selfconsistently
the dynamics of geometry and matter \cite{ZPM}.

In this phenomenological approach we ignore microscopic details and
concentrate on most relevant macroscopic magnitudes like particle number
density, temperature, entropy production. The net effect of the interaction
term between both fields is collected in the particle number rate of change,
and the finite temperature takes into account the backreaction effect of the
produced particles. A phenomenological approach was initially proposed in
\cite{ZPM}. The methods used in that paper for the solution of the evolution
equations lead to a vanishing bulk viscous contribution to the speed of sound.
However their model depends strongly on this contribution so that their
procedure of calculation needs to be improved. In this direction we introduce
a physically founded perturbative scheme that yields more satisfactory
results.

The plan of this paper is as follows. Section 2 introduces the two-fluid
magnitudes and section 3 presents main equations of the model in the
one-temperature picture. In section 4 we introduce an expansion in powers of
a parameter that measures de departure from perfect fluid behavior. This
allows us to solve the system of equations order by order. In section 5 the
model with a linear ansatz for the decay rate is solved, and we show that it
reaches the reheating temperature only for very large times. We propose in
section 6 a quadratic ansatz for the decay rate that is better motivated both
from microphysical and geometrical point of view. We investigate thoroughly
the evolution of the physical magnitudes and we find that the behavior of this
model is very interesting. Finally, conclusions are stated in section 7.


\section{Two-fluid picture}

For the two-fluid model we assume that
the energy-momentum tensor $T ^{ik}$ splits into  two perfect fluid parts,

\begin{equation}
T ^{ik} = T ^{ik}_{_{1 }}
+ T ^{ik}_{_{2 }} {\mbox{ , }}
\label{1}
\end{equation}

\noindent
with ($A = 1, 2$)

\begin{equation}
T^{ik}_{_{A }} = \rho_{_{A }} u^{i}u^{k} 
+ p_{_{A }} h^{ik}{\mbox{ .}} \label{2}
\end{equation}

\noindent $\rho_{_{A }}$ is the energy density and $p_{_{A }}$ is the
equilibrium pressure of species $A$. For simplicity we assume that both
components share the same $4$-velocity $u^{i}$. The quantity $h^{ik}$ is the
projection tensor $h^{ik} =g^{ik} + u^{i}u^{k}$. The particle flow vector
$N_{_{A }}^{i}$ of species $A$ is defined as

\begin{equation}
N_{_{A }}^{i} = n_{_{A }}u^{i}{\mbox{ , }} \label{3}
\end{equation}

\noindent where $n _{_{A }}$ is the particle number density. We are interested
in situations where neither the particle numbers nor the energy-momenta of the
components are separately conserved. The
balance laws for the particle numbers are

\begin{equation}
N _{_{A } ;i}^{i} = \dot{n}_{_{A }}
+ \Theta n_{_{A }} = n _{_{A }} \Gamma _{_{A}}
{\mbox{ , }}
\label{4}
\end{equation}

\noindent where $\Theta \equiv u^{i}_{;i}$ is the fluid expansion and $\Gamma
_{_{A}}$ is the rate of change of the number of particles of species $A$.
There is particle production for $\Gamma _{_{A}} > 0$ and particle decay for
$\Gamma _{_{A}} < 0$, respectively. For $\Gamma _{_{A}} = 0$, we have separate
particle number conservation.

Interactions between the fluid components amount to the mutual exchange of
energy and momentum. Consequently, there will be no local energy-momentum
conservation for the subsystems separately. Only the energy-momentum tensor of
the system as a whole is conserved.

Denoting the loss- and source-terms in the separate balances by $t ^{i}_{_{A
}}$, we write

\begin{equation}
T ^{ik}_{_{A } ;k} = - t _{_{A }}^{i} {\mbox{ , }}
\label{9}
\end{equation}

\noindent
implying

\begin{equation}
\dot{\rho}_{_{A }}
+ \Theta\left(\rho_{_{A }}
+ p_{_{A }}\right)
= u _{a} t _{_{A }}^{a}
{\mbox{ , }}
\label{10}
\end{equation}

\noindent
and

\begin{equation}
\left(\rho _{_{A }}
+ p _{_{A}}\right) \dot{u}^{a}
+ p _{_{A } ,k}h ^{ak}
= - h ^{a}_{i}t ^{i}_{_{A }} {\mbox{ .}}
\label{11}
\end{equation}

\noindent All the considerations to follow will be independent of the specific
structure of the $t _{_{A }}^{i}$. 

Each component is governed by a separate Gibbs equation

\begin{equation}
T _{_{A }} \mbox{d} s _{_{A }}
= \mbox{d} \frac{\rho _{_{A }}}{n _{_{A }}}
+ p _{_{A }}
\mbox{d} \frac{1}{n _{_{A }}} {\mbox{ , }}
\label{12}
\end{equation}

\noindent where $s _{_{A }}$ is the entropy per particle of species $A$. Using
eqs.(\ref{4}) and (\ref{10}) one finds for the time behaviour of the entropy
per particle

\begin{equation}
n _{_{A }} T _{_{A }}
\dot{s}_{_{A }} = u _{a} t _{_{A }}^{a}
- \left(\rho _{_{A }}
+ p _{_{A }}\right) \Gamma _{_{A}} {\mbox{ .}}
\label{13}
\end{equation}

\noindent With nonvanishing source terms in the balances for $n _{_{A }}$ and
$\rho _{_{A }}$, the change in the entropy per particle is different from zero
in general. 

The equations of state of the fluid components are assumed to have the general
form

\begin{equation}
p_{_{A }} = p_{_{A }}
\left(n_{_{A }}, T_{_{A }}\right) \label{14}
\end{equation}

\noindent
and

\begin{equation}
\rho_{_{A }} =
\rho_{_{A }} \left(n_{_{A }},
T_{_{A }}\right){\mbox{ , }}
\label{15}
\end{equation}

\noindent i.e., particle number densities $n_{_{A }}$ and temperatures $T_{_{A
}}$ are regarded as the  basic thermodynamical variables. The temperatures of
the fluids are different in general.

Differentiating  relation (\ref{15}), using the balances (\ref{4}) and
(\ref{10}) as well as the general relation

\begin{equation}
\frac{\partial \rho_{_{A }}}{\partial n_{_{A }}} = 
\frac{\rho_{_{A }} + p_{_{A }}}
{n_{_{A }}} 
- \frac{T_{_{A }}}{n_{_{A }}}
\frac{\partial p_{_{A }}}
{\partial T_{_{A }}} {\mbox{ , }}
\label{16}
\end{equation}

\noindent that follows from the requirement that the entropy is a state
function, we find the following expression for the temperature behaviour
\cite{Calv,LiGer,ZPRD}:

\begin{equation}
\dot{T}_{_{A }}  = - T_{_{A}} \left(\Theta -
\Gamma _{_{A}} \right)
\frac{\partial p_{_{A}}/\partial T_{_{A}}}{\partial \rho_{_{A}}/
\partial T_{_{A}}}
+ \frac{u _{a} t _{_{A}}^{a} - \Gamma _{_{A}} \left(\rho _{_{A}}
+ p _{_{A}}\right)}
{\partial \rho _{_{A}}/ \partial T _{_{A}}}
{\mbox{ .}}
\label{17}
\end{equation}

The entropy flow vector $S _{_{A}}^{a}$ is defined by

\begin{equation}
S _{_{A}}^{a} = n _{_{A}} s _{_{A}} u ^{a} {\mbox{ , }}
\label{18}
\end{equation}

\noindent and the contribution of component $A$ to the entropy production
density becomes

\begin{eqnarray}
S _{_{A} ;a}^{a} &=& n _{_{A}} s _{_{A}} \Gamma _{_{A}} + n _{_{A}} 
\dot{s}_{_{A}}\nonumber\\
&=& \left(s _{_{A}} - \frac{\rho _{_{A}}
+ p _{_{A}}}{n _{_{A}}T _{_{A}}}
\right)
n _{_{A}}\Gamma _{_{A}} + \frac{u _{a} t _{_{A}}^{a}}{T _{_{A}}}
{\mbox{ , }}
\label{19}
\end{eqnarray}

\noindent
where relation (\ref{13}) has been used.

According to eq.(\ref{9}) the condition of energy-momentum conservation for
the system as a whole,

\begin{equation}
\left(T _{_{1}}^{ik} + T _{_{2}}^{ik}\right)_{;k} = 0 {\mbox{ , }}
\label{20}
\end{equation}

\noindent
implies

\begin{equation}
t _{_{1}}^{a} = - t _{_{2}}^{a} {\mbox{ .}}
\label{21}
\end{equation}

\noindent There is no corresponding condition, however, for the particle
number balance as a whole. Defining the integral particle number density $n$
as

\begin{equation}
n = n _{_{1}} + n _{_{2}} {\mbox{ , }}
\label{22}
\end{equation}

\noindent
we have

\begin{equation}
\dot{n} + \Theta n = n \Gamma {\mbox{ , }}
\label{23}
\end{equation}

\noindent
where

\begin{equation}
n \Gamma = n _{_{1}}\Gamma _{_{1}} + n _{_{2}}\Gamma _{_{2}}
{\mbox{ .}}
\label{24}
\end{equation}

\noindent $\Gamma$ is the rate by which the total particle number $n$ changes.

The entropy per particle is \cite{Groot}

\begin{equation}
s _{_{A}} = \frac{\rho _{_{A}} + p _{_{A}}}{n _{_{A}}
T _{_{A}}} - \frac{\mu _{_{A}}}
{T _{_{A}}}
{\mbox{ , }}
\label{25}
\end{equation}

\noindent where $\mu _{_{A}}$ is the chemical potential of species $A$.
Introducing the last expression into eq.(\ref{19}) yields

\begin{equation}
S ^{a}_{_{A} ;a} = - \frac{\mu _{_{A}}}{T _{_{A}}}
n _{_{A}}\Gamma _{_{A}}
+ \frac{u _{a}t ^{a}_{_{A}}}{T _{_{A}}} {\mbox{ .}}
\label{26}
\end{equation}
For the total entropy production density

\begin{equation}
S ^{a}_{;a} = S ^{a}_{_{1} ;a} + S ^{a}_{_{2} ;a}
\label{27}
\end{equation}

\noindent
we obtain

\begin{equation}
S ^{a}_{;a} = - \frac{\mu _{_{2}}}{T _{_{2}}}n \Gamma
- \left(\frac{\mu _{_{1}}}{T _{_{1}}} - \frac{\mu _{_{2}}}{T _{_{2}}}
\right)
n _{_{1}}\Gamma _{_{1}}
+ \left(\frac{1}{T _{_{1}}} - \frac{1}{T _{_{2}}}\right)u _{a}
t _{_{1}}^{a} {\mbox{ .}}
\label{28}
\end{equation}

We assume that the entropy per particle of
each of the components is preserved. The particles decay or are produced
with a fixed entropy $s _{_{A}}$. This `isentropy' condition amounts to the
assumption that the particles at any stage are amenable to a perfect fluid
description. When  $\dot{s}_{_{A}} = 0$, we get from (\ref{13})

\begin{equation}
u _{a} t ^{a}_{_{A}} = \left(\rho _{_{A}} + p _{_{A}}\right)
\Gamma _{_{A}}
{\mbox{ , }}
\label{30}
\end{equation}

\noindent
and combining eqs.(\ref{21}) and (\ref{30}), one has

\begin{equation}
u _{a} t ^{a}_{_{1}} = \left(\rho _{_{1}} + p _{_{1}}\right)
\Gamma _{_{1}}
= - u _{a} t ^{a}_{_{2}} = - \left(\rho _{_{2}} + p _{_{2}}\right)
\Gamma _{_{2}}
{\mbox{ , }}
\label{31}
\end{equation}

\noindent
which provides us with a relation between the rates
$\Gamma _{_{1}}$ and
$\Gamma _{_{2}}$:
\begin{equation}
\Gamma _{_{2}} = - \frac{\rho _{_{1}} + p _{_{1}}}
{\rho _{_{2}} + p _{_{2}}}
\Gamma _{_{1}}
{\mbox{ .}}
\label{32}
\end{equation}

\noindent Inserting the last relation into equation (\ref{24}) yields

\begin{equation}
n \Gamma = n _{_{1}}\Gamma _{_{1}} h _{_{1}}
\left(\frac{1}{h _{_{1}}}
- \frac{1}{h _{_{2}}}\right) {\mbox{ .}}
\label{33}
\end{equation}

\noindent The quantities $h _{_{A}} \equiv \left(\rho _{_{A}} + p _{_{A}}
\right)/ n _{_{A}}$ are the enthalpies per particle. 

With the relations (\ref{30}) and (\ref{33}) the entropy production 
density (\ref{28}) becomes

\begin{equation}
S ^{a}_{;a} = \left(\rho _{_{1}} + p _{_{1}}\right)
\left[\frac{n _{_{1}} s _{_{1}}}{\rho _{_{1}} + p _{_{1}}}
- \frac{n _{_{2}} s _{_{2}}}{\rho _{_{2}} + p _{_{2}}}\right]
\Gamma _{_{1}}
= n _{_{1}}\Gamma _{_{1}} h _{_{1}}
\left[\frac{s _{_{1}}}{h _{_{1}} }
- \frac{ s _{_{2}}}{h _{_{2}}}\right]
{\mbox{ .}}
\label{34}
\end{equation}

\noindent We emphasize that according to the equations of state (\ref{14}) and
(\ref{15}) the quantities $\rho _{1}$, $p _{_{1}}$ and $s _{_{1}}$ depend on
$T _{_{1}}$, while $\rho _{_{2}}$, $p _{_{2}}$ and $s _{_{2}}$ depend on $T
_{_{2}}$. In general, we have $T _{_{1}} \neq T _{_{2}}$.

To complete the two-fluid model we give the thermodynamical properties of both
fluids. If we ignore effects associated with particle creation, after
inflation the field $\phi$ oscillates near the point $\phi=0$ at a frequency
given by its mass $m$. Its oscillation amplitude $\Phi$ falls off as $t^{-1}$
and its mean energy density $\rho_{\phi}= m^2\Phi^2/2$ decreases as $a^{-3}$,
in the same way as pressureless dust \cite{Bel85}. Hence we represent the
massive inflaton field as a fluid, named $1$, of nonrelativistic particles
with mass $m$, energy density $\rho_1=\rho_\phi$, pressure $p_1\ll \rho_1$ and
number density $n_1\simeq\rho_1/m$. On the other hand, the decay products of
the scalar field $\phi$ are ultrarelativistic for $m \gg m_\chi$, and we model
them as an ideal relativistic fluid of massless particles, with energy density
$\rho_2$, equilibrium pressure $p_2=\rho_2/3$ and particle density $n_2$. We
also assume that both fluids are classical and ideal and that each species
remain in thermal equilibrium along the reheating process. That is, we neglect
dissipative effects for each of the fluids in comparison with particle
production effects.


\section{Effective one-temperature picture}

We have also the effective one-temperature
alternative description of a two-component fluid that is
based on a single Gibbs-equation for the system as a whole:

\begin{equation}
T \mbox{d}s = \mbox{d} \frac{\rho}{n} + p \mbox{d} \frac{1}{n}
- \left(\mu _{_{1}} - \mu _{_{2}}\right) \mbox{d} \frac{n _{_{1}}}{n} 
{\mbox{ , }}
\label{35}
\end{equation}

\noindent where $p$ is the equilibrium pressure, $\rho$ is the energy density
and $s$ is the entropy per particle. The temperature $T$ is the equilibrium
temperature of the whole system and is defined by \cite{UI,Zim96a}

\begin{equation}
\rho_{_{1}}\left(n_{_{1}},T_{_{1}}\right)
+ \rho_{_{2}}\left(n_{_{2}},T_{_{2}}\right) 
= \rho \left(n, n _{_{1}}, T\right)
{\mbox{ .}}
\label{38}
\end{equation}

\noindent
Furthermore, we assume that the cosmic fluid as a whole is
characterized by the equations of state

\begin{equation}
p = p\left(n, n _{_{1}}, T\right)
\label{36}
\end{equation}

\noindent
and

\begin{equation}
\rho = \rho \left(n, n _{_{1}}, T\right){\mbox{ , }}
\label{37}
\end{equation}

The temperatures $T _{_{1}}$ and $T _{_{2}}$ do not appear as variables in the
effective one-temperature description. An (approximate) equilibrium
for the entire system is assumed to be established through the interactions
between the subsystems on the right-hand side of the balances (\ref{10}) and
(\ref{11}). 
In this picture the cosmic
fluid splits into two effective fluid components whose equations of state are
assumed to have the same form as those of the "real" fluids, but shear this
common temperature. So, these two effective fluids are in a (formal) thermal
equilibrium between them while the "real" fluids are not.

If we set $T _{_{1}}=T _{_{2}}=T$
in (\ref{25}) and make the splitting for the energy density of the system

\begin{equation} \label{rho}
\rho \left(T\right) = \rho _{_{1}}\left(T\right) + \rho
_{_{2}}\left(T\right)
\end{equation}

\noindent
and for its equilibrium pressure

\begin{equation}\label{p}
p\left(T\right) = p _{_{1}}\left(T\right)
+ p _{_{2}}\left(T\right),
\end{equation}

\noindent
the description based on relation
(\ref{35}) is consistent with the description relying on the relations
(\ref{12}) for $ns \left(T\right) = n _{_{1}} s _{_{1}}\left(T\right) + n
_{_{2}} s _{_{2}}\left(T\right)$.
As long as the pressures are
those for classical gases, $p _{_{A}} = n _{_{A}} T$, the equilibrium
pressure $p$ of the system as a whole depends on $n = n _{_{1}} + n _{_{2}}$
only and the separate dependence on $n _{_{1}}$  on the right-hand side of
eq.(\ref{36}) may be omitted.

For components out of equilibrium, even when $\Gamma =
\Gamma _{_{1}} = \Gamma _{_{2}} = 0$, the total pressure is generally
different to the sum of the partial pressures. In this case we have
\cite{Zim96a}

\begin{equation}
p_{_{1}}\left(n_{_{1}},T_{_{1}}\right)
+ p_{_{2}}\left(n_{_{2}},T_{_{2}}\right) 
= p\left(n, T\right)+\pi_d
{\mbox{ .}} \label{39}
\end{equation}

\noindent instead of (\ref{p}). Here $\pi_d$ is the viscous pressure arising
from differential temperature variation rate between both fluids
\cite{Zim96a}. From eq.(\ref{17}) the cooling rate $\dot{T}_{_{1}}/T _{_{1}}$
is different from $\dot{T}_{_{2}}/T _{_{2}}$ even for $\Gamma _{_{1}} = \Gamma
_{_{2}} = 0$ if the subsystems are governed by different equations of state.
The expansion of the Universe tends to increase the difference between $T
_{_{1}}$ and $T _{_{2}}$.

On the other hand, deviations from detailed balance, i.e., $\Gamma
_{_{A}} \neq 0$, leading to $\Gamma \neq 0$ in general,  generate an
effective `reactive' bulk pressure $\pi$. For the corresponding
energy-momentum tensor of the system as a whole we write

\begin{equation}
T ^{ik} = \rho u ^{i}u ^{k} + \left(p + \pi \right) h ^{ik}
{\mbox{ .}}
\label{40}
\end{equation}

\noindent The reactive bulk pressure is determined by the consistency of the
expression for the entropy production density obtained in the two-temperature
picture (\ref{34}) and the expression obtained in the one-temperature picture

\begin{equation} \label{dS}
S^\mu_{;\mu}= s n \Gamma + n\dot s
\end{equation}

\noindent The last one arises from the entropy flow vector $S^\mu=snu^\mu$
\cite{Isr76}, where $s$ is the entropy per particle and $u^\mu$ is the
four-velocity of the fluid. It was estimated that $\pi_d$ is one order of
magnitude smaller than $\pi$ \cite{Zim97b}.

From the Gibbs-equation (\ref{35}) one finds for the change
in the entropy per particle

\begin{equation}
n \dot{s} = - \frac{\Theta }{T}\pi - \frac{\rho +
p}{T}\Gamma
- \frac{n _{_{1}} n _{_{2}}}{n}\left(\frac{\mu _{_{1}}
- \mu _{_{2}}}{T}\right)
\left(\Gamma _{_{1}} - \Gamma _{_{2}}\right) {\mbox{ .}}
\label{42}
\end{equation}

\noindent
Even for $\dot{s}_{_{1}} = \dot{s}_{_{2}} = 0$ we have $\dot{s} \neq 0$ in
general. However, during the preheating stage, particle production dominates
over per particle entropy change as the main entropy production source.

The equilibrium temperature of the system $T$ is the key magnitude of our
model. On the one hand we identified it with the reheating temperature, and on the
other hand its definition (\ref{38}) provides a mean to account for
backreaction effects of the created relativistic particles on their massive
counterparts. Its evolution law is given by \cite{Zim97b}

\begin{equation}\label{dT}
\frac{\dot{T}}{T} =  \left(\Gamma-\Theta\right)
\frac{\partial p/\partial T}{\partial{\rho }/\partial T}
- \frac{\Theta \pi }{T \partial\rho/\partial T} \,
\end{equation}

The Einstein field equations for a spatially-flat
Robertson-Walker spacetime are

\begin{equation} \label{ZPM28}
3H^{2}  = \kappa\rho
\ ,  
\end{equation}

\begin{equation}
\dot{H}  = - {\kappa\over 2}\left(\rho + p + \pi\right)
\ , \label{ZPM29}
\end{equation}

\noindent where $\kappa$ is Einstein's gravitational constant and $H=
\dot{a}/a$, $a$ is the cosmic scale factor. We use units such that $c=1$,
$k_B=1$ and $\hbar=1$, then $\kappa=8\pi/M_P^2$, where $M_P$ is the Planck
mass.  Finally a dot denotes derivative with respect to comoving time.
Conservation of (\ref{40}) leads to

\begin{equation} \label{drho}
\dot\rho+3H\left(\gamma\rho+\pi\right)=0
\end{equation}

\noindent
where

\begin{equation} \label{gamma}
\gamma(t) \equiv 1 + {p(t)\over\rho(t)}
\end{equation}

\noindent is the time dependent polytropic index. Using (\ref{ZPM28}) and
(\ref{ZPM29}) we get

\begin{equation} \label{pi}
\kappa\pi=-3\gamma H^2-2\dot H
\end{equation}

When $m \gg T $, the equations of state for the nonrelativistic fluid
$1$ are:

\begin{equation}
\rho _{1} = n _{1} m + {\textstyle{3\over2}}
n _{1} T \ , \quad
p _{1} = n _{1} T
\label{ZPM36}
\end{equation}

\noindent
while for the relativistic fluid $2$ are:

\begin{equation}
\rho _{2} = 3 n _{2}T \ ,\quad
p _{2} = n _{2} T
\label{ZPM37}
\end{equation}

\noindent
For these fluids we have

\begin{equation} \label{gamma2}
\gamma=1+\frac{nT}{n_1m+\left(\frac{3}{2}n_1+3n_2\right)T}
\end{equation}

To calculate the evolution of the temperature we need also
$\Gamma$. Using (\ref{33}) and the equations of state (\ref{ZPM36}) and
(\ref{ZPM37}), we obtain

\begin{equation} \label{Gamma}
\Gamma=\frac{n_1 m}{4nT}Q
\end{equation}

\noindent where the decay rate $Q=|\Gamma_1|$ is an input to the model that
must be chosen from a fundamental microphysical theory or from
phenomenological considerations.

\section{The quasiperfect expansion}

\noindent We assume that the viscous
effects, are small. If $\tau$ is the mean
interaction time of the particles of the fluid, we have that $\nu=\left(\tau
H\right)^{-1}$ is the number of interactions in an expansion time.
Perfect fluid behavior occurs in the limit $\nu\to\infty$. Small departures from
this behavior occur for large $\nu$, and a consistent hydrodynamical description
of the fluids requires $\nu>1$. Thus we are lead to assume that $\tau H$ is
small and we propose the following "quasiperfect" expansion in powers of
$\nu^{-1}$

\begin{equation} \label{H1a}
H=H_0\left(1+\frac{h_1}{\nu} +\cdots\right)
\end{equation}

\begin{equation} \label{pi1}
\pi=\frac{\pi_1}{\nu}+\frac{\pi_2}{\nu^2}+\cdots
\end{equation}

\noindent
where $\pi_i$, will be fixed by the thermodynamical theory adopted.
Inserting (\ref{H1a}) and (\ref{pi1}) in (\ref{pi}), we find the equations that
determine the coefficients of the expansion up to first order in
$\nu^{-1}$

\begin{equation} \label{dH0}
\dot H_0+(3/2)\gamma H_0^2=0
\end{equation}

\begin{equation} \label{dH1}
\dot h_1=-\frac{\pi_1}{2H_0}
\end{equation}

\noindent
where we have assumed that $\dot\tau\ll\tau H$.
Solving these equations we get

\begin{equation} \label{H0}
H_0(t)=\frac{2}{3\int dt \gamma(t)}
\end{equation}

\noindent
and

\begin{equation} \label{H2}
H=H_0\left(1-\frac{\kappa}{2\nu}\int dt\, \frac{\pi_1}{H_0}
+\cdots\right)
\end{equation}

In the particular case we choose the truncated transport equation of Causal
Irreversible Thermodynamics for the bulk
viscosity pressure

\begin{equation}
\pi + \tau\dot{\pi}  =  - 3\zeta H .
\label{ZPM25}
\end{equation}

\noindent where we identify $\tau$ with  the relaxation time,
 the bulk viscosity coefficient $\zeta$ is given by \cite{Maar96}

\begin{equation}
\frac{\zeta}{\tau} =  c_b^2  (\rho + p),
\label{ZPM26}
\end{equation}

\noindent
$c_b$ is the bulk viscous contribution to the speed
of sound $v$, $v^2=c_s^2+c_b^2\leq 1$, and
$c_s$ is the adiabatic sound speed, we find

\begin{equation} \label{pi2}
\kappa\pi\approx-\frac{9}{\nu} c_b^2 \gamma H_0^2
\end{equation}

\noindent
Also it can be easily seen  that $H_0$ is an exact solution of equations
(\ref{ZPM28}), (\ref{ZPM29}), (\ref{gamma}), (\ref{ZPM25}) and (\ref{ZPM26})
provided $c_b=0$.

To get an estimation of the physical parameters in the quasiperfect regime it
is enough to keep calculations at zero order in $\nu^{-1}$ and we can neglect
the viscous terms in (\ref{drho}), (\ref{dT}) and (\ref{gamma2}). In this order of
approximation the results are independent of the form of the transport
equation. Nevertheless this approach will allow us to give a reasonable
description of the reheating process.
Then (\ref{drho}) becomes

\begin{equation} \label{drho0}
\dot \rho+3H_0 \gamma\rho\approx 0
\end{equation}

\noindent
whose solution is

\begin{equation} \label{rho0}
\rho(t)\approx \frac{\rho_0}{\left(\int dt\gamma(t)\right)^2}
\end{equation}

\noindent
where $\rho_0$ is a positive integration constant. Also (\ref{dT}) becomes

\begin{equation} \label{dT0}
\frac{\dot T}{T}\approx
\frac{2n}{3\left(n_1+2n_2\right)}
\left(\frac{n_1m}{4nT}Q-3H_0\right)
\end{equation}

\noindent
Then, when the decay rate $Q$ is large enough,
it may overcompensate the adiabatic term during the initial `preheating'
stage and make the temperature rise.

\section{Linear reheating}

Following \cite{ZPM} we assume first a linear ansatz $Q=\beta H$, with an
adimensional constant $\beta>0$.

To solve the coupled system of equations (\ref{23}), (\ref{4}),
(\ref{dT0}) and (\ref{gamma2})
for the evolution for  $n$, $n_1$,
$n_2$, $T$ and $\gamma$, we use an iterative scheme that starts from a fully
nonrelativistic stage: $n_1=n$, $\gamma=1$. In this regime

\begin{equation} \label{H1}
H_0\approx \frac{2}{3t}
\end{equation}

\noindent
Then (\ref{dT0}) reduces to

\begin{equation} \label{dT1}
\frac{\dot T}{T}\simeq
\frac{4}{9t}
\left(\frac{m\beta}{4T}-3\right)
\end{equation}

\noindent
whose solution is

\begin{equation} \label{T1}
T(t)=T_1\left[1-\left(\frac{t_0}{t}\right)^{4/3}\right]
\end{equation}

\noindent where $t_0$ is an arbitrary integration constant with dimension of
time, and
$T_1=m\beta/12$. In order to describe the reheating scenario we need to choose
$t_0>0$. As in the previous inflationary stage the Universe "supercools"
\cite{Kolb90}, we assume that the inflation period ends at $t\approx t_0$ when
 $T\approx 0$, and we take solution (\ref{T1}) for $t>t_0$. Then the
temperature grows monotonically and approaches asymptotically its maximum
value $T_1$,  which we could call the reheating temperature of the linear
model, for $t\to\infty$.
We see from (\ref{T1}) that the temperature rises quite slowly. It takes a
time over $30 t_0$ for the temperature to approach at $1\%$ of $T_1$. For
this reason, we will not pursue further the consequences of the linear ansatz
and in a following section we will propose an improved ansatz for $Q$ which
leads a much faster rise in temperature.

\section{Quadratic reheating}

In the previous section, we have seen that a linear ansatz for $Q$ leads to a
very slow process of reheating. A microscopic interaction term like
$g\phi^2\chi^2$, suggests an effective quadratic decay rate $\Gamma\sim
\Phi^2\sim t^{-2}$. Here we look for an improved ansatz that takes into
account the scalar nature of $\Gamma$. In this direction we propose
$Q=\bar\beta R$, where $R=6\left(\dot H+2 H^2\right)$ is the curvature scalar
and $\bar\beta$ is a positive constant with dimension of time.

At zero order in $\nu^{-1}$ we obtain

\begin{equation} \label{Q2}
Q=6\bar\beta\left(2-\frac{3\gamma}{2}\right)H_0^2\equiv 6\tau_1 H_0^2
\end{equation}

\noindent
This expression shows that $Q\ge 0$ for $1\le\gamma\le 4/3$ and vanishes in a
radiation dominated era, when $\Phi$
is very small.  Inserting (\ref{Q2}) in (\ref{Gamma})
and using (\ref{dT0}) we obtain at zero order in $\nu^{-1}$

\begin{equation} \label{dT2}
\frac{\dot T}{T}\simeq\frac{nH_0}{n_1+2n_2}\left(\frac{n_1m\tau_1}{nT}H_0-2\right)
\end{equation}

\noindent
and we solve it using the iterative scheme.

\subsection{Nonrelativistic regime}

In this regime when the fluid is dominated by massive particles,
equation (\ref{dT2}) becomes

\begin{equation} \label{dT3}
\dot T+2H_0 T\simeq m\tau_1 H_0^2
\end{equation}

\noindent
whose general solution is

\begin{equation} \label{T4}
T(t)=\frac{4m\tau_1}{3t}\left[1-\left(\frac{t_0}{t}\right)^{1/3}\right]
\end{equation}

\noindent As in the linear model, the reheating scenario corresponds to
$t_0>0$, and we take the solution (\ref{T4}) for $t>t_0$. This solution also
starts from $T=0$, but this time it rises violently to a maximum temperature
of reheating

\begin{equation} \label{Tr}
T_r=\frac{9\tau_1}{64 t_0}m
\end{equation}

\noindent in a period $t_r-t_0=37 t_0/27$, where $t_r$ is the time when
maximum temperature occurs. Assuming that inflation ends at $t_0=\delta/ m$,
where $\delta$ is a numeric constant, this rising period is $1.37 \delta/m$.
This is a time of the same order as the period of oscillation of the inflaton
field $2\pi/m$. After that, the temperature begins to fall. Though these
calculations are carried out in the approximation  $T\ll m$, and this
implies that ( \ref{Tr} ) is strictly valid only for $\tau_1\ll t_0$, it also
suggests that $T_r$ may be of order $m$ when $\tau_1$ is large enough.

Inserting (\ref{H1}) and (\ref{Q2}) in (\ref{4}) and solving the resulting
equation for the number density of massive particles we find

\begin{equation} \label{n_1}
n_1(t)\simeq n_{10} \frac{e^{8\tau_1/\left(3t\right)}}{t^2}
\end{equation}

\noindent where $n_{10}$ is a positive integration constant, so that $n_1(t)$ is a
decreasing function.

Using (\ref{23}), (\ref{Gamma}), (\ref{T4}) and (\ref{n_1}), we find the
total number particle density in the first departure from the nonrelativistic
initial state

\begin{equation} \label{n}
n(t)\simeq\frac{n_0}{t^2}\left[1+\frac{n_{10}}{2n_0}\int dx
\frac{e^{8\tau_1/\left(3t_0 x\right)}}{x-x^{2/3}}\right]
\end{equation}

\noindent
where $x=t/t_0$, and $n_0$ is a positive constant. Assuming that the
reheating temperature is much smaller than $m$ we obtain the approximated
expression

\begin{equation} \label{n2}
n(t)\simeq \frac{n_0}{t^2}\left[1+\frac{3n_{10}}{2n_0}
\ln\left(x^{1/3}-1\right)\right]
\end{equation}

The expression (\ref{n_1}) for the density of nonrelativistic particles is a
decreasing function, while the expression (\ref{n2}) for the total number
density has a peak. The consistency requirement $n_1\leq n$ is satisfied after
the time $t_p$ when $n_1(t_p)=n(t_p)\equiv n_p$. This time is given by

\begin{equation} \label{xp}
x_p=\left[1+\exp\left(\frac{2\left(\alpha-1\right)}{3\alpha}\right)\right]^3
\end{equation}

\noindent
where $\alpha=n_{10}/n_0$, and
marks in our model the starting point of the preheating stage.
We demand the temperature

\begin{equation} \label{Tp}
T(t_p) \approx 4\left(\frac{4}{3}\right)^3 \left(x_p^{1/3}-1\right) T_r
\end{equation}

\noindent to be very low and this occurs when $x_p$ is close to $1$, so
that $\alpha\ll 1$. This condition corresponds to an explosive
production of light bosons. In effect, from 
( \ref{22} ), ( \ref{n2} ) and ( \ref{n_1} ) we get\begin{equation} \label{n_2}
n_2(t)\approx \frac{3 n_{10}}{2t^2}\ln\left(
\frac{x^{1/3}-1}{x_p^{1/3}-1}\right)
\end{equation}

\noindent and just after $t_p$ this particle density grows linearly with an
exponentially large slope

\begin{equation} \label{n_22}
n_2(t)\approx \frac{n_{10} e^{2/\left(3\alpha\right)}}{2t_0^2}
\left(x-x_p\right), \qquad x-x_p\ll 1
\end{equation}

\noindent reaching a large peak density $n_p/\alpha$ in a time $x_m\approx
1+3\alpha/4$. After that, both $n$ and $n_2$ begin to fall rapidly as the
effect of expansion dominates over particle production.

In the regime of large particle production
we can neglect in (\ref{dS}) the change in $s$, so that
the entropy production density is approximately given by

\begin{equation}
S ^{a}_{;a} \approx sn\Gamma=\frac{sn_{10}\,
e^{8\tau_1/\left(3t\right)}}
{2t^2\left(t-t_0^{1/3}t^{2/3}\right)}
\label{S}
\end{equation}

\noindent
and we obtain the following lower bound for the entropy density

\begin{equation} \label{S1}
S >\frac{s n_{10}}{2t_0^2}\int_{x_p}^{x_r}
\frac{dx}{x^2\left(x-x^{2/3}\right)} \approx
-\frac{3s n_{10}}{2t_0^2}\ln\left(x_p^{1/3}-1\right)
\approx \frac{sn_0}{t_0^2}
\end{equation}

\noindent
Thus we find that the generated entropy per massive particle
$S/n(t_p)>s/\alpha$ is very high.

\subsection{Intermediate regime}

Here $\rho_1\approx\rho_2$, i.e. $ n_1 m \approx 3 n_2 T$ so that $n_2\gg n_1$
and $\gamma\approx{7\over6}$. At the microscopic level this corresponds to the
regime when the backreaction of the produced particles ceases to be
negligible. We can estimate the time $t_i$ when this regime begins from the
equation $\rho(t_i)=2\rho_1(t_i)$, using (\ref{ZPM36}), (\ref{rho0}),
(\ref{T4}) and (\ref{n_1}). Therefore $t_i$ becomes a function of $K\equiv
m\tau_1/\delta$ and the intermediate regime is reached provided $K>3\ln 2/8$.
For lower values of $K$ particle production rate is not large enough to
compensate the faster decrease of relativistic matter energy density due to
cosmic expansion. On the other hand, $K$ cannot be much larger than this lower
bound if $T_r\ll m$. For instance, we get $x_i\approx 41$ and $T_r/m\approx
0.038$ for $K=0.27$ . We
show in Fig. 1 the evolution of $\rho_1$ and $\rho_2$ for this value
of $K$, from fully nonrelativistic to intermediate regime.

In this intermediate regime holds

\begin{equation}
H_0(t) \approx  \frac{4}{7t}
\label{H10}
\end{equation}

\begin{equation} \label{T10}
T(t)\approx T_0^{(i)}\frac{e^{-24\tau_1/\left(49 t\right)}}{t^{4/7}}
\end{equation}

\begin{equation} \label{n_110}
n_1(t)\approx n_{10}^{\left(i\right)}
\frac{e^{96\tau_1/\left(49 t\right)}}{t^{12/7}}
\end{equation}

\begin{equation} \label{n10}
n(t)\approx n_0^{\left(i\right)}
\frac{e^{-72\tau_1/\left(49 t\right)}}{t^{12/7}}
\end{equation}

\noindent where $T_0^{(i)}$, $n_{10}^{(i)}$ and $n_0^{(i)}$ are positive
integration constants fixed by matching at $t_i$ expressions (\ref{T10}),
(\ref{n_110}) and (\ref{n10}) with their nonrelativistic counterparts
(\ref{T4}), (\ref{n_1}) and (\ref{n}). After the initial stage of large
ultrarelativistic particle creation in the nonrelativistic regime the
nonrelativistic particle decay process slows down. All particle densities and
the temperature decrease, though the dilution and cooling rates are smaller
because of the slower cosmic expansion.  The whole evolution of the
temperature and particle number densities along the nonrelativistic and
intermediate regimes is plotted in Figs. 2-4 for $K=0.27$.

\subsection{Ultra-relativistic regime}

In this last stage of reheating, the energy density
becomes dominated by
the radiation fluid, i.e. $n_2T \gg n_1m$, 
$\gamma \approx {4\over3}$.
Therefore  we have

\begin{equation}
H_0 \approx  \frac{1}{2t}
\label{H20}
\end{equation}

\begin{equation} \label{T20}
T(t)\approx\frac{T_0^{\left(u\right)}}{t^{1/2}}
\end{equation}

\begin{equation} \label{n20}
n(t)\approx\frac{n_0^{\left(u\right)}}{t^{3/2}},
\qquad n_1(t)\approx\frac{n_{10}^{\left(u\right)}}{t^{3/2}}
\end{equation}

\noindent where $T_0^{(u)}$, $n_{10}^{(u)}$ and $n_0^{(u)}$ are positive
integration constants. This stage marks the end of reheating and the cosmic
medium approaches a perfect, relativistic fluid with vanishing viscosity and
particle production. The temperature and particle densities continue falling
at about the adiabatic rate. This is smaller than the intermediate stage rate
as the radiation-dominated universe expands even slower. A small remnant of
nonrelativistic particles may remain.

\section{Conclusions}

With the aim of understanding the process of reheating we have developed a
phenomenological model of two interacting fluids. In this model the
oscillating inflaton field is modeled after a nonrelativistic fluid of massive
particles that decay into an ultrarelativistic fluid of massless particles.

We have carried out a ``quasiperfect'' perturbative expansion to solve the
Einstein equations. This corresponds to a small viscous pressure regime. To
the lowest order in this expansion, the thermodynamical magnitudes do not
depend on the transport equation for the bulk viscosity.

A particle decay rate, linear in the expansion rate, does not yield a suitable
model as the maximum reheating temperature is reached only for very long
times. However an alternative quadratic ansatz for the decay rate gives a good
behavior of the thermodynamical magnitudes. The last one was investigated in
detail along its evolution from the initial nonrelativistic stage to the final
domination by ultrarelativistic particles. The equilibrium temperature of the
whole system and the produced particle number density rise violently in a time
comparable to the oscillation period of the inflaton field. Also, large
amounts of entropy are produced in this initial stage of reheating. The time
when the energy density of the decay products becomes comparable with the
nonrelativistic particles, is similar to the microscopic calculations for the
time when backreaction effects become important.

In a future paper we will investigate phenomenological reheating models far
from the perfect fluid regime.

\noindent
{\bf Acknowledgements}\\
This work has been partially supported by the Universidad de Buenos Aires
under Grant TX-93.

\newpage

\noindent
{\Large \bf Figure Captions}

\bigskip \noindent Figure 1. Plot of the equilibrium energy densities of
nonrelativistic and relativistic particles, $\rho_1$ and $\rho_2$ in units
of $\rho_0=\rho(t_0)$ vs. adimensional time $x$, from $x=1$ to
$x\simeq x_i$, for $K=0.27$. In this case $x_i\simeq 41$.

\bigskip \noindent Figure 2. Plot of the equilibrium temperature $T$ in units
of inflaton mass $m$ vs. adimensional time $x$  along the
nonrelativistic and intermediate regimes, for $K=0.27$.

\bigskip \noindent Figure 3. Plot of the nonrelativistic particle number
density $n_1$ in units of the initial density $n_p$ vs. adimensional time $x$
along the nonrelativistic and intermediate regimes, for $K=0.27$.

\bigskip \noindent Figure 4. Plot of the relativistic particle number density
$n_2$ in units of the initial density $n_p$ vs. adimensional time $x$ along
the nonrelativistic and intermediate regimes, for $\alpha=0.1$ and $K=0.27$.
The inserted plot shows a detail of the evolution of $n_2$ about its peak.

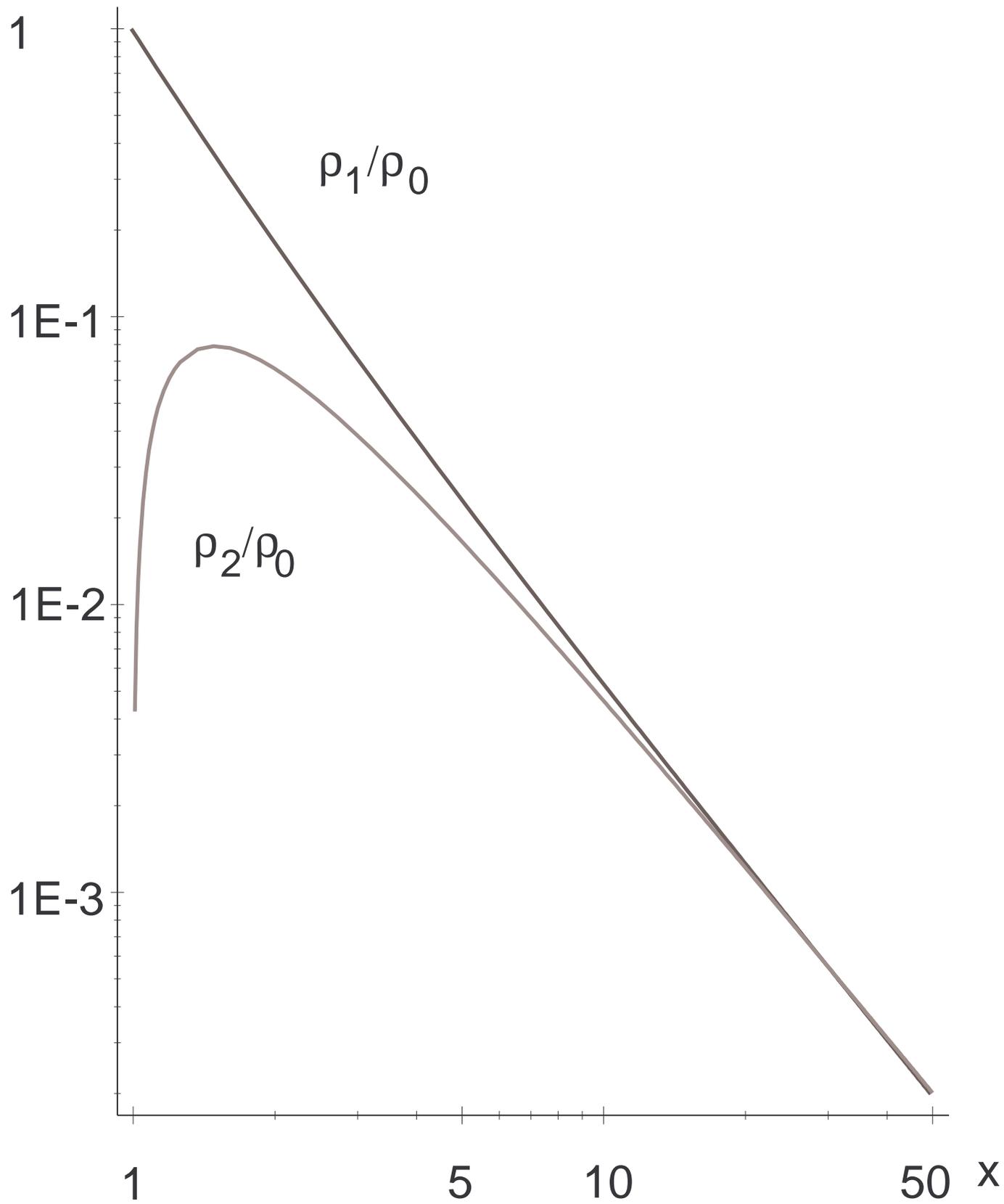
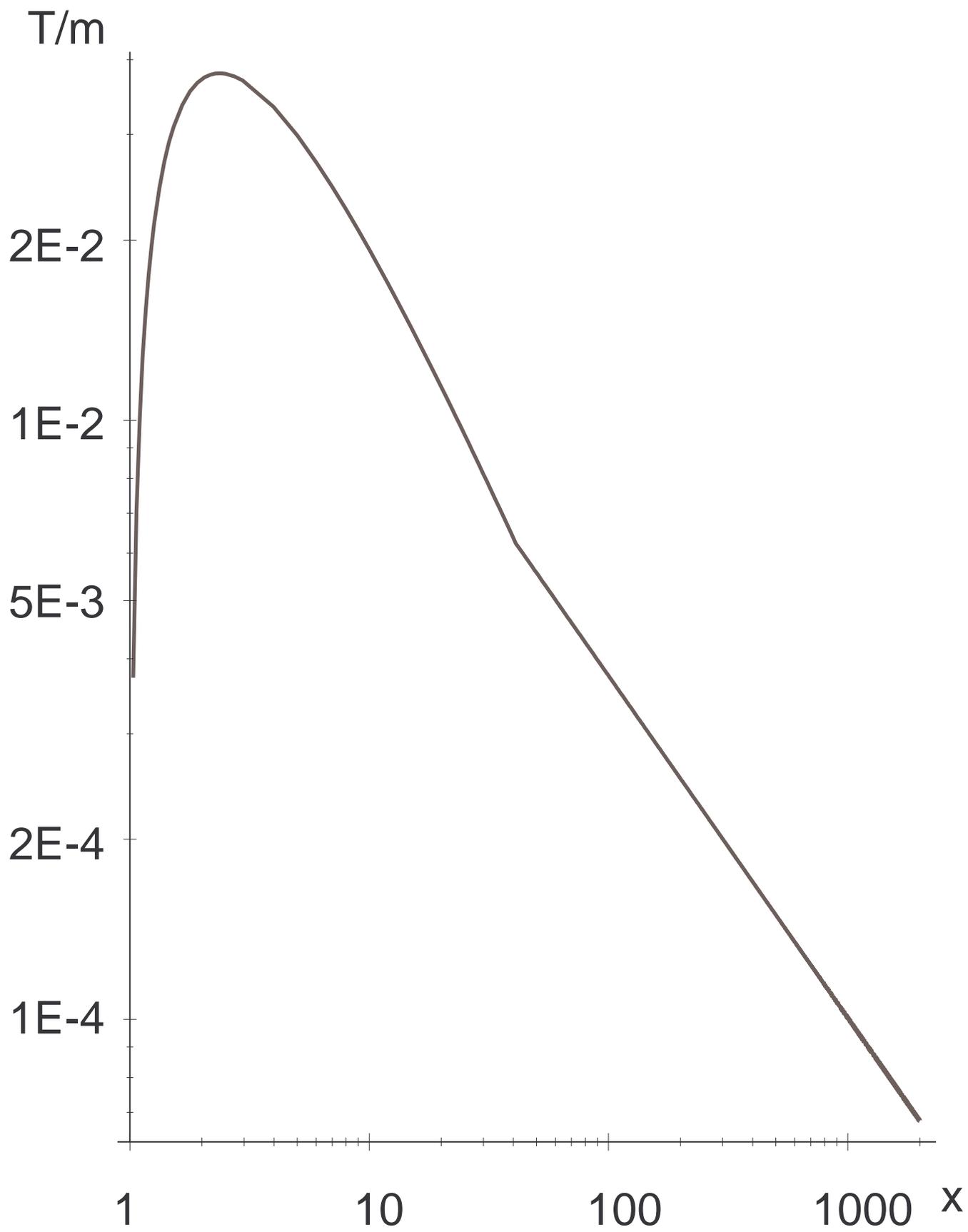
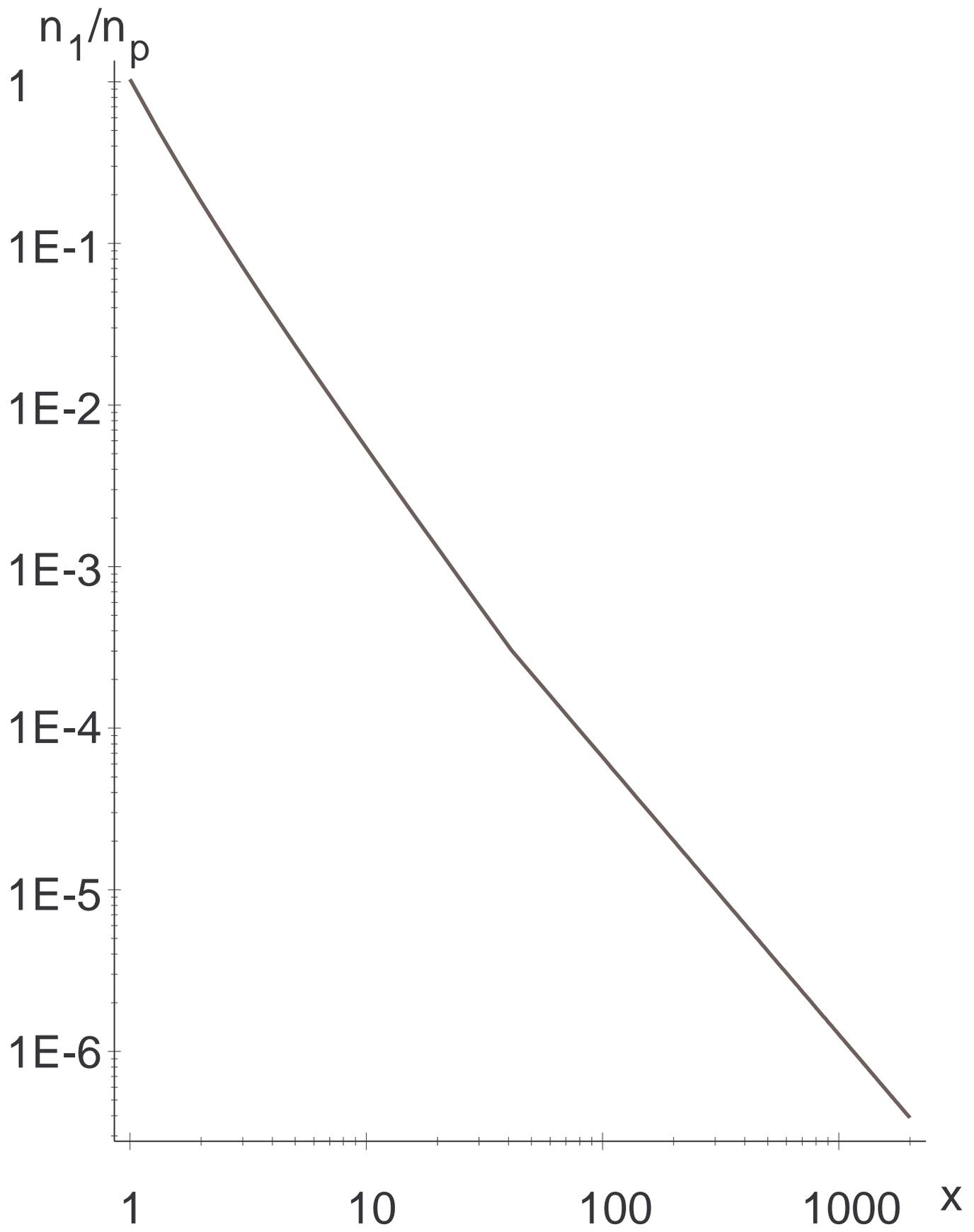
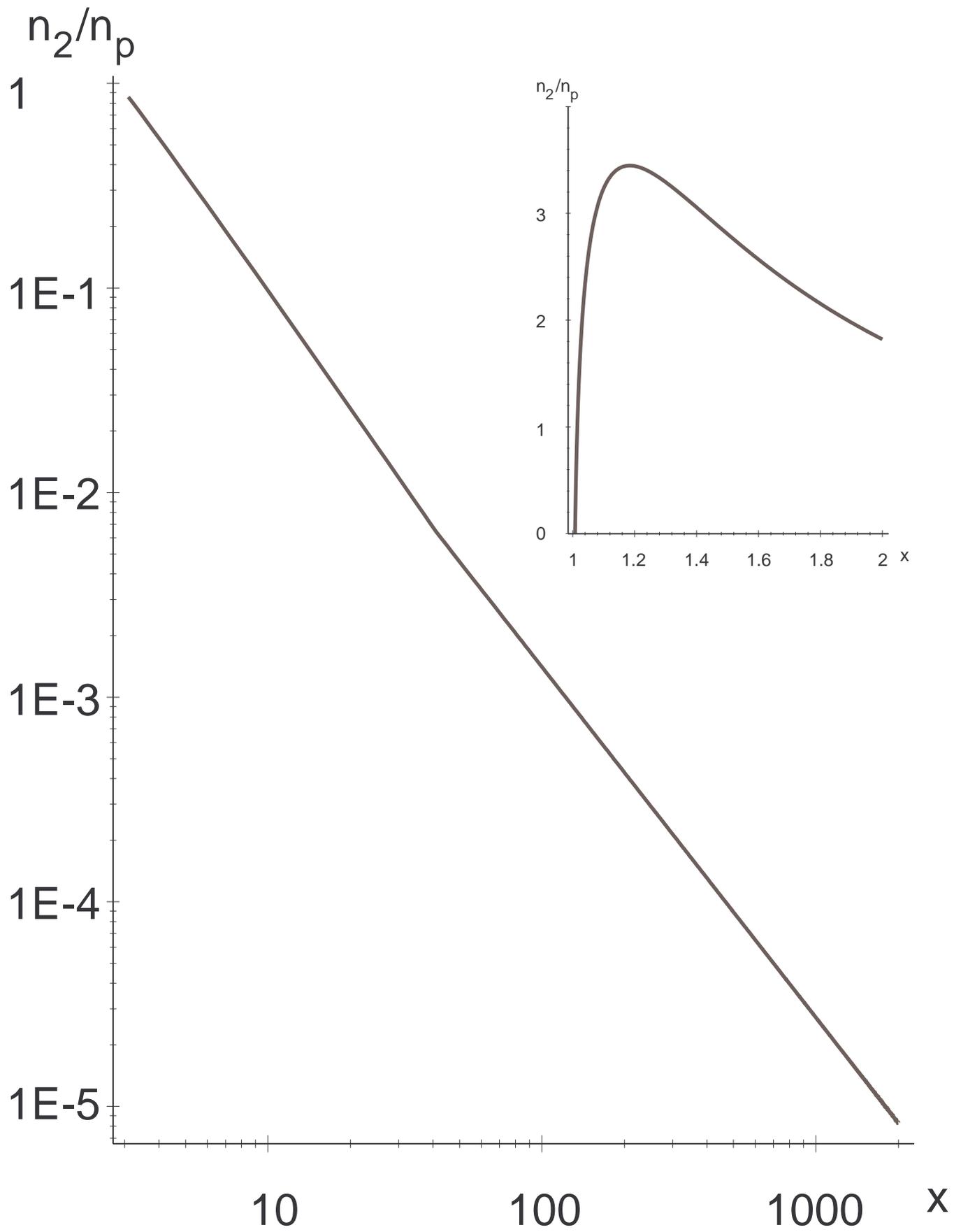
\end{document}